\documentclass[12pt]{article}
\usepackage{a4wide}

\usepackage{graphicx}

\newcommand{\eps}{\varepsilon}
\newcommand{\pfrac}[2]{\left(\frac{#1}{#2}\right)}

\newcommand{\bbbone}{\hbox{\rm 1\kern-3pt l}}
\newcommand{\slp}{p\kern-5pt/}

\newcommand{\dalem}{\hbox{\,\vbox{\hrule height0.2pt\hbox{\vrule width0.2pt
  \hbox{\vbox to5pt{\hbox to5pt{\hfill}\vfill}}\vrule width.6pt%
  \kern-0.1pt}\hrule height0.6pt}\,}}

\begin{document}

\begin{center}
{\Large\bf Dynamical generation of electroweak scale from the
  conformal sector: {\it A strongly coupled Higgs via the Dyson--Schwinger
  approach}}\\[1cm]
{\Large Arpan Chatterjee$^1$, Marco Frasca$^2$,\\[7pt]
Anish Ghoshal$^3$ and Stefan Groote$^1$}\\[12pt]
$^1$ F\"u\"usika Instituut, Tartu Ulikool,
W.~Ostwaldi 1, EE-50411 Tartu, Estonia\\[7pt]
$^2$ Rome, Italy\\[7pt]
$^3$ Institute of Theoretical Physics, Faculty of Physics,\\
University of Warsaw, ul.\ Pasteura 5, 02-093 Warsaw, Poland
\end{center}
\vspace{1cm}
\begin{abstract}\noindent
We propose a novel pathway to generate the electroweak (EW) scale via
non-pertur\-bative dynamics of a conformally invariant scalar sector at the
classical level. We provide a method to estimate the non-perturbative EW scale
generation using the exact solution of the background equations of motion in
a scalar theory via the Dyson--Schwinger approach. Particularly, we find an
analytical result for the Higgs mass in the strongly coupled regime in terms
of its quartic self interaction term and the cut-off scale of the theory. We
also show that the Higgs sector is an essential part of the Standard Model as,
without it, a Yang--Mills gauge theory cannot acquire mass even if a
self-interaction term is present. Our analysis lead to a more realistic model
building with possible solutions to the gauge hierarchy problem and, in
general, to the dynamical generation of any scales scales in nature, be it
the visible sector or the dark sector.
\end{abstract}

\newpage

Long time ago, Coleman and Weinberg postulated a dynamical generation of
gauge symmetry breaking that can be realized via radiative symmetry breaking
arising due to quantum corrections generic in any quantum field theory (QFT
)~\cite{Coleman:1973jx}.  However, when the mechanism is applied to the
Standard Model (SM) gauge theories, the masses of the gauge bosons are found
to be greater than that of the Higgs boson, $m_{Z,W} > m_H$, which is
experimentally disfavoured. Nonetheless, realistic dynamical generations of
the electroweak (EW) scale can be achieved and have been explored extensively
in the framework of Beyond the SM (BSM) approaches by several
authors~\cite{Adler:1982ri,Salvio:2014soa,Einhorn:2014gfa,Einhorn:2016mws,%
Einhorn:2015lzy,Englert:2013gz,Holthausen:2009uc,Meissner:2006zh,Foot:2007as,%
Farzinnia:2013pga}. Moreover, in the context of non-minimally coupled gravity,
scale invariant models naturally consist of flat inflationary
potentials~\cite{Khoze:2013uia,Kannike:2014mia,Rinaldi:2014gha,Salvio:2014soa,%
Kannike:2015apa,Kannike:2015fom,Barrie:2016rnv,Tambalo:2016eqr}, and the mass
scale of dark matter can be dynamically generated with
ease~\cite{Hambye:2013sna,Karam:2015jta,Kannike:2015apa,Kannike:2016bny,%
Karam:2016rsz}. Therefore, classically conformal theories have always been
seen as a direction of model building towards the dynamically generated mass
scale as a possible resolution to the gauge hierarchy problem in the
SM~\cite{Foot:2007iy,AlexanderNunneley:2010nw,Englert:2013gz,Hambye:2013sna,%
Farzinnia:2013pga,Altmannshofer:2014vra,Holthausen:2013ota,Salvio:2014soa,%
Einhorn:2014gfa,Kannike:2015apa,Farzinnia:2015fka,Kannike:2016bny}. See
Refs.~\cite{1306.2329,1410.1817,0902.4050,0909.0128,1210.2848,1703.10924,%
1807.11490} for other studies of conformal invariance and dimensional
transmutation of energy scales~\cite{2012.11608,1812.01441}. However, all
these approaches either involve further assumptions on the weak perturbation
theory or assume the involvement of some additional BSM dark sector apart from
just the SM of particle physics.

As basic element of the EW sector, the SM Higgs mechanism for SM gauge symmetry
breaking is formulated by postulating a scalar potential, consisting of a
mass-like contribution to the Lagrangian proportional to the square of the
field, and a self-interaction part of the power of four (self-quartic) in the
scalar field. While the latter is essential for the mechanism to be possible,
the former is set merely by hand, with a negative value of the coefficient
allowing for the mechanism to emerge. In this short note we show that gauge
theory itself bears the possibility that such a parameter with negative value
can evolve from the solution of a mass gap equation, breaking the conformal
invariance down to the Lagrangian of the Standard Model. We will show that
such a mass term could be dynamically generated without the need to lose scale
invariance as a fundamental symmetry of the theory at the classical level.
Besides, we show that the effect of breaking the electroweak symmetry is a
dynamical effect by itself.

Our considerations outlined in this note are based on the exact solution of
the background equations of motion in Yang--Mills theory following the analytic
approach of Dyson--Schwinger equations, originally devised by Bender, Milton
and Savage in Refs.~\cite{Bender:1999ek}. Due to the possible feature of the
fact that the Green's functions of the theory can be represented analytically,
we can understand the effect of the background on the interactions that
remains valid even in the strongly-coupled regime~\cite{Frasca:2015yva}. This
mathematical tool has been widely devised and has found several applications,
ranging from  QCD~\cite{Frasca:2021yuu,Frasca:2021mhi,Frasca:2022lwp,%
Frasca:2022pjf,Chaichian:2018cyv} to the scalar
sector~\cite{Frasca:2015wva}, as well as to extensions to other types of
models including the gauge sector and string-inspired non-local
theories~\cite{Frasca:2019ysi,Chaichian:2018cyv,Frasca:2017slg,Frasca:2016sky,%
Frasca:2015yva,Frasca:2015wva,Frasca:2013tma,Frasca:2012ne,Frasca:2009bc,%
Frasca:2010ce,Frasca:2008tg,Frasca:2009yp,Frasca:2008zp,Frasca:2007uz,%
Frasca:2006yx,Frasca:2005sx,Frasca:2005mv,Frasca:2005fs}. As an application to
particle physics phenomenology, some of the authors explored non-perturbative
hadronic contributions to the muon anomalous magnetic moment
$(g-2)_\mu$~\cite{Frasca:2021yuu}, QCD in the non-perturbative
regime~\cite{Frasca:2021mhi,Frasca:2022lwp,Frasca:2022pjf}, Higgs-Yukawa
theory~\cite{Frasca:2023qii}, finite temperature field
theory~\cite{Frasca:2023eoj}, and in early universe cosmology like
non-perturbative false vacuum decay and phase
transitions~\cite{Frasca:2022kfy,Calcagni:2022tls,Calcagni:2022gac}, dark
energy~\cite{Frasca:2022vvp}, and explorations of the mass gap and confinement
in string-inspired infinite-derivative and higher-derivative Lee--Wick
theories motivated by UV-completion of gravity~\cite{Frasca:2020jbe,%
Frasca:2020ojd,Frasca:2021iip,Frasca:2022duz,Frasca:2022gdz}.

\begin{figure}
\centering
\includegraphics[width=0.8\textwidth]{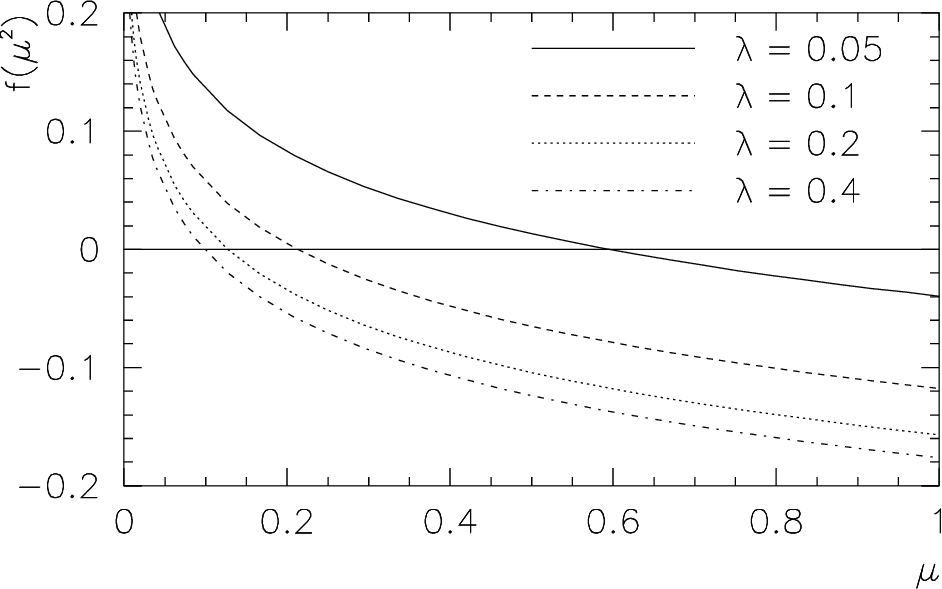}
\caption{\it For a given value of $\lambda$, the function $f(\mu^2)$, related
to the mass gap, always obtains a zero for a value of the renormalisation
scale $\mu$ chosen large enough.\label{fig1}}
\end{figure}

To start with, we consider a scalar field $\phi$ with a classically conformal
invariant Lagrangian
\begin{equation}
L=\frac12(\partial\phi)^2-\frac\lambda4\phi^4.
\end{equation}
where $\lambda$ represents the self-interaction quartic coupling constant. We
solve this theory exactly through the solution of the tower of Dyson--Schwinger
equations, showing that this theory matches the well-known Higgs sector of the
SM. Indeed, from Ref.~\cite{Frasca:2015yva} we obtain the following set of
Dyson--Schwinger equations:
\begin{eqnarray}
\label{eq:Gn}
\lefteqn{\partial^2G_1(x)+\lambda\Big(\left(G_1(x)\right)^3}\nonumber\\&&
  +3G_2(x,x)G_1(x)+G_3(x,x,x)\Big)\ =\ 0,\nonumber\\[7pt]
\lefteqn{\partial^2G_2(x,y)+\lambda\Big(3\left(G_1(x)\right)^2G_2(x,y)}
  \nonumber\\&&
  +3G_3(x,x,y)G_1(x)+3G_2(x,x)G_2(x,y)\nonumber\\&&
  +G_4(x,x,x,y)\Big)\ =\ -i\delta^4(x-y),\nonumber\\[7pt]
\lefteqn{\partial^2G_3(x,y,x,z)+\lambda\Big(6G_1(x)G_2(x,y)G_2(x,z)}
  \nonumber\\&&
  +3G_1^2(x)G_3(x,y,z)+3G_2(x,z)G_3(x,y,x)\nonumber\\&&
  +3G_2(x,y)G_3(x,x,z)+3G_2(x,x)G_3(x,y,z)\nonumber\\&&
  +3G_1(x)G_4(x,x,y,z)+G_5(x,x,x,y,z)\Big)\ =\ 0,\nonumber\\[7pt]
\lefteqn{\partial^2G_4(x,y,z,w)+\lambda\Big(6G_2(x,y)G_2(x,z)G_2(x,w)}
  \nonumber\\&&
  +6G_1(x)G_2(x,y)G_3(x,z,w)+6G_1(x)G_2(x,z)G_3(x,y,w)\nonumber\\&&
  +6G_1(x)G_2(x,w)G_3(x,y,z)+3G_1^2(x)G_4(x,y,z,w)\nonumber\\&&
  +3G_2(x,y)G_4(x,x,z,w)+3G_2(x,z)G_4(x,x,y,w)\nonumber\\&&
  +3G_2(x,w)G_4(x,x,y,z)+3G_2(x,x)G_4(x,y,z,w)\nonumber\\&&
  +3G_1(x)G_5(x,x,y,z,w)+G_6(x,x,x,y,z,w)\Big)\ =\ 0,\nonumber\\&&
  \ldots
\end{eqnarray}
where $G_n(x_1,\ldots,x_n)$ are the correlation functions of the theory. As we
know from experimental evidences, the Standard Model is translation invariant.
This means that a constant is the only possible choice for $G_1$. Thus, we
assume $G_1(x)=v$ with a constant $v$ as the unique solution for the one-point
correlation function. Besides, $G_2(x,x)$ can be interpreted as a mass term
generated by quantum corrections and is a constant as well. This constant is
infinite and will need regularization. However, we can write such a mass term
as
\begin{equation}
\label{eq:mu2}
\kappa=3\lambda G_2(x,x)=3\lambda\int\frac{d^4p}{(2\pi)^4}G_2(p).
\end{equation}
The constant solution $G_1(x)=v$ is obtained by solving the equation
\begin{equation}\label{veq}
\kappa v+\lambda v^3=0,
\end{equation}
where the unstable case $v=0$ must be excluded. From Eq.~(\ref{eq:Gn}), for
$G_2(x,y)$ we have
\begin{equation}
\partial^2G_2(x,y)+3\lambda v^2G_2(x,y)+\kappa G_2(x,y)=-i\delta^4(x-y),
\end{equation}
and by using Eq.~(\ref{veq}) we recover the Higgs boson mass $m_H^2=-2\kappa$.
Thus, from Eq.~(\ref{eq:mu2}) we obtain
\begin{equation}
\kappa=3\lambda\int\frac{d^4p}{(2\pi)^4}G_2(p)
  =3\lambda\int\frac{d^4p}{(2\pi)^4}\frac{i}{p^2+2\kappa+i\eta},
\end{equation}
and a Wick rotation yields
\begin{eqnarray}\label{eq:mu}
\kappa=3\lambda\int\frac{d^Dp_E}{(2\pi)^D}\frac1{p_E^2-2\kappa}
  =\frac{3\lambda\Gamma(\eps-1)}{(4\pi)^{2-\eps}}(-2\kappa)^{1-\eps}
  =\frac{6\lambda\kappa}{(4\pi)^2}\left[\frac1\eps+1
  -\ln\pfrac{-2\kappa}{\mu^2}\right]
\end{eqnarray}
in dimensional regularisation ($D=4-2\eps$) with the renormalisation scale
$\mu=\mu_{\overline{\rm MS}}$ of the $\overline{\rm MS}$ scheme. At this
stage, we notice an essential difference to the Coleman--Weinberg mechanism,
i.e., by breaking conformal invariance, we perfectly mimic the Higgs mechanism.
No problems of any kind arises for the observed mass spectrum of the SM. It is
easy to check that a (sufficiently large) value of $\lambda$ a solution can be
found such that $\kappa<0$. In order to see this, we consider the gap equation
$f(\mu^2)=0$ arising from Eq.~(\ref{eq:mu}) by subtracting the singularity and
setting $m_H^2=-2\kappa$, where
\begin{equation}
f(\mu^2)=\frac{m_H^2}{2\lambda}-\frac{6m_H^2}{(4\pi)^2}
  \left[1-\ln\pfrac{m_H^2}{\mu^2}\right].
\end{equation}
The function $f(\mu^2)$ is plotted in Fig.~\ref{fig1} for different values of
$\lambda$. By this gap equation, the Higgs mass is fixed to
$m_H^2=\mu^2\exp(1-4\pi^2/3\lambda)$. The plot in Fig.~\ref{fig1} shows how
the value of $\lambda$ determines possible values of $\mu$ for which the
criterion $m_H^2>0$ leads to conformal symmetry breaking and the generation of
the EW scale, i.e., the point where a single Higgs boson emerges in our
approach.

In order to see how this simple model maps on the Higgs sector of the Standard
Model, we consider the electroweak Lagrangian~\cite{Schwartz:2014sze}
\begin{eqnarray}
{\cal L}_H&=&\left|\left(i\partial_\mu+\frac{g_2}2W_\mu^a\sigma_a
  +\frac{g_1}2B_\mu\right)H\right|^2-\frac\lambda2\left(H^\dagger H\right)^2
  \nonumber\\&&
  -\bar Q_LY_u\tilde Hu_R-\bar Q_LY_dHd_R-\bar L_LY_eHe_R+\mathrm{h.c.}\qquad
\end{eqnarray}
with a charge conjugated Higgs field $\tilde H$, where $W$ and $B$ are the
gauge fields, $Q_L$ and $L_L$ are built up by three generations of left handed
SM quark and lepton fields, respectively,  $f_R$ ($f=u,d,e$) contain three
generations of right handed quark or lepton fields, $Y_f$ are the matrices of
Yukawa couplings, $g_2$ and $g_1$ are the gauge couplings for the group
$SU(2)\times U(1)$, $\sigma_a$ are the Pauli matrices, and $\lambda$ is the
self-coupling of the Higgs field
\begin{equation}
H=\frac1{\sqrt2}\left(
\begin{array}{c}
h_1 + ih_2 \\ h_3 + ih_4
\end{array}
\right)
\end{equation}
that is a doublet under the gauge group $SU(2)$ with real valued components
$h_i$ ($i=1,\ldots,4$). The physical Higgs particle is given by the real part
of the state with weak hypercharge $y=1$ and weak isospin $t_3=-1/2$. As we
have shown, the term $|H|^2$ is generated by the dynamics. Indeed, we will
obtain the Dyson--Schwinger equation
\begin{eqnarray}
\lefteqn{\partial^2G_1(x)+\lambda\Big(G_2(x,x)G_1(x)+G_1^\dagger(x)G_2(x,x)}
  \\&&
  +G_1^\dagger(x)F_2(x,x)+G_1(x)|G_1(x)|^2+G_3^\dagger(x,x,x)\Big)=0.\nonumber
\end{eqnarray}
We can consistently take the component $G_{1h_3}(x)$ as the only one taking a
mass value $G_{2h_3}(x,x)$ while we can choose the other constant $F_2(x,x)$
to be zero through renormalization. Therefore, our argument about conformal
invariance for the scalar field given in the beginning applies also to the
Higgs sector in the Standard Model. This can be seen by observing that the
general equation for the one-point (1P) correlation function of the Higgs
field can be solved by perturbation theory, taking at the leading order the
symmetry breaking solution. What concerns fermions, based on the partition
function of the SM that is not given explicitly here but necessary to evaluate
averages in quantum field theory, one has to evaluate the averages for the
correlation functions of the equations of motion like
\begin{eqnarray}
\langle\gamma^\mu\left(i\partial_\mu+\frac{g_2}2W_\mu^a\sigma_s
  +\frac{g_1}2B_\mu\right)Q_L\rangle=\langle Y_u\tilde Hu_R+Y_dHd_R+j\rangle,
\end{eqnarray}
where $j$ represents all the contributions coming from the Lagrangian of the
SM due to the interaction of the fermion field $\psi$ with the other fields in
the theory. In a straightforward way, one observes that a mass term is
generated as $G_1=v$ like
\begin{equation}
\langle Y_u\tilde Hu_R\rangle=vY_u\langle u_R\rangle,\qquad
\langle Y_d Hd_R\rangle=vY_d\langle d_R\rangle.
\end{equation}

Note that the Higgs sector is essential in the SM as, without it, a gauge
theory cannot acquire mass even if a self-interaction term is present, in
case that translation invariance holds. Therefore, in order to complete our
proof, we show that if we insist on the conservation of translation invariance
as a fact experimentally observed, the only way to generate masses in the SM
is through a scalar sector. Thus, we have to prove that the gauge sector
itself does not show a mass gap. For this aim, we solve the Yang--Mills theory
with gauge group $SU(N)$ (this convenience can easily be generalized to other
gauge groups) based on the Lagrangian
\begin{equation}
{\cal L}_{\rm YM}=-\frac14F_{\mu\nu}^aF^{\mu\nu}_a
\end{equation}
by considering the Dyson--Schwinger equations~\cite{Bender:1999ek}. Solving
these in Feynman gauge, we obtain (cf.\ Appendix~B)~\cite{Chaichian:2018cyv}
\begin{eqnarray}\label{DSE}
\lefteqn{\partial^2G_{1\nu}^{a}(x)
  +gf^{abc}\Big(\partial^\mu G_{2\mu\nu}^{bc}(x,x)
  +G_{1\mu}^{b}(x)\partial^\mu G_{1\nu}^{c}(x)}\nonumber\\&&\quad
  -\partial_\nu G_{2\mu}^{\mu bc}(x,x)
  -G_{1\mu}^{b}(x)\partial_\nu G_{1}^{\mu c}(x)\Big)\nonumber\\&&
  +gf^{abc}\partial^\mu\Big(G_{2\mu\nu}^{bc}(x,x)+G_{1\mu}^b(x)G_{1\nu}^c(x)
  \Big)\nonumber\\&&
  +g^2f^{abc}f^{cde}\Big(G_{3\mu\nu}^{\mu bde}(x,x,x)\nonumber\\&&\quad
  +G_{2\mu}^{\mu bd}(x,x)G_{1\nu}^{e}(x)
  +G_{2\nu}^{\mu be}(x,x)G_{1\mu}^{d}(x)\nonumber\\&&\quad
  +G_{2\mu\nu}^{de}(x,x)G_{1}^{\mu b}(x)
  +G_{1}^{\mu b}(x)G_{1\mu}^{d}(x)G_{1\nu}^{e}(x)\Big)\nonumber\\
  &=&gf^{abc}\partial_\nu\Big(P^{cb}_2(x,x)
  +\bar P^{b}_1(x)P^{c}_1(x))\Big)+j_\nu^a(x),\\[7pt]&&
  \partial^2 P^{a}_1(x)+gf^{abc}\partial^\mu(K^{bc}_{2\mu}(x,x)
  +P^{b}_1(x)G_{1\mu}^{c}(x))\ =\ 0,\nonumber
  \end{eqnarray}
where $G^a_{1\mu}(x)$ is the 1P-correlation function for the gauge field,
$G_{2\mu\nu}^{ab}(x,y)$ is the two-point (2P) correlation function for the
gauge field, and $P_1^a(x)$ and $P_2^{ab}(x,y)$ are 1P- and 2P-correlation
functions for the ghost field. Contained in the second equation is a mixed
2P-correlation function $K_2^{ab}(x,y)$ between the gauge and the ghost
fields. Higher-order correlation functions appear in this equation evaluated
at the same points. In general, these are infinite constants to be
renormalized. Note that the ghost propagator can be decoupled by taking
$P_1^a(x)=0$ that, according to the second equation, results in
$K^{bc}_{2\mu}(x,x)$ being a constant. Dimensionally regularised, the solution
$P_2^{ab}(x,y)\propto\delta_{ab}/|x-y|^2$ leads again to
$P_2^{ab}(x,x)\rightarrow 0$.

As we have shown in our preceding works~\cite{Frasca:2015yva,%
Chaichian:2018cyv,Frasca:2019ysi,Frasca:2017slg,Frasca:2020ojd,Frasca:2021iip},
the Dyson--Schwinger equation for the 1P-correlation function $G_{1\mu}^a(x)$
can be solved exactly, and inserting this solution into the Dyson--Schwinger
equation for the 2P-correlation function, a mass gap becomes manifest in the
mass-like term $G_{2\mu\nu}^{ab}(x,x)$, at the same time restoring the
translation invariance. We have recognised that this solution is consistent
with lattice data~\cite{Bogolubsky:2007ud,Cucchieri:2007md,Oliveira:2007px}.
This solution works well in the IR limit. On the other hand, if we impose
translation invariance from the very beginning, the choice
$G_{1\mu}^a(x)=n_\mu w^a$ for the 1P-correlation function with $n^\mu n_\mu=1$
being a Minkowski vector and $w^a$ a gauge group constant grants translation
invariance. In this case, we are left with the algebraic equation
\begin{eqnarray}
\lefteqn{f^{abc}f^{cde}\Big(G_{3\mu\nu}^{\mu bde}(x,x,x)}\nonumber\\&&
+\delta_{bd}G_{2\mu\nu}(x,x)G_{1}^{\mu e}(x)
+\delta_{eb}G_{2\nu\rho}(x,x)G_{1}^{\rho d}(x)\nonumber\\&&
+\delta_{de}G_{2\mu\nu}(x,x)G_{1}^{\mu b}(x)
+G_{1}^{\mu b}(x)G_{1\mu}^{d}(x)G_{1\nu}^{e}(x)\Big)\nonumber\\
&=&f^{abc}f^{cde}G_{3\mu\nu}^{\mu bde}(x,x,x)
+f^{abc}f^{cbe}G_{2\mu\nu}(x,x)n^\mu w^e\\&&
+f^{abc}f^{cdb}G_{2\nu\rho}(x,x)n^\rho w^d
+f^{abc}f^{cde} w^b w^d w^e n_\nu\ =\ 0,\nonumber
\end{eqnarray}
where we used the mapping $G_{2\mu\nu}^{ab}(x,y)=\delta_{ab}G_{2\mu\nu}(x,y)$.
A group argument, based on the fact that
$G_{3\mu\nu}^{\mu bde}(x,y,z)=f^{bde}G_{3\mu\nu}^\mu(x,y,z)$ and
$f^{bde}f^{cde}\sim\delta_{bc}$, grants that the three-point correlation
function does not contribute. Therefore, we are left only with a trivial
solution $w^a=0$. This does not yield a mass gap to the theory. A similar
study was carried out in Ref.~\cite{Frasca:2015wva} where the choice of the
background solution broke translation invariance. However, in the case studied
in this note we are able to reproduce exactly the well-known scalar sector of
the SM generating dynamically the odd term $-m^2$ that, in the textbook
formulation, breaks conformal symmetry without breaking translation invariance.

In summary, we have investigated a generic Higgs field that is conformally
invariant at the classical level. This field receives a vacuum expectation
value (vev) due to non-perturbative dynamics involving the self-quartic term.
Following a novel technique developed by Bender \textit{et al.}, we were able
to compute this vev analytically in Eqn.~(\ref{eq:mu}). Such strongly coupled
non-perturbative dynamics lead to an effective scalar potential of the Higgs
field that mimics the SM Higgs potential with a negative mass squared term.
We have shown that the potential arising exhibits proper minima that can give
rise to electroweak scale breaking (see Fig.~\ref{fig1}).

With dynamically generated scales due to a strongly coupled scalar sector, our
results shed light on and are generally applicable to a wide range of
approaches for realistic BSM model building. For example, one may envisage a
classically conformal SM $\times$ SU(2)$_D$ model, where the Higgs vev
that breaks the SU(2)$_D$ symmetry generation dynamically may have its origin
in non-perturbative scenarios. In general, such symmetry breaking scales can
be very high compared to the EW scale and may involve very interesting dark
matter phenomenology~\cite{Holthausen:2009uc,Hambye:2013dgv,Antipin:2014qva,%
Barman:2021lot,Frasca:2024pmv}.

As an outlook, starting from classically scale-invariant theories and scale
generation via non-perturbative dynamics, we allude to a dynamical
explanation for the generation of (any) scales in nature, and subscribe to the
notion that no scales are special in nature. In the end, this might include
also such fundamental scales in nature as the EW scale, the seesaw scale, or
the Planck scale. Therefore, the approach presented here provide a possibly
intriguing avenue to understand why different kinds of fundamental
interactions (for example, gravity and EW) are vastly different in their
strengths. However, a detailed generation of the Planck and EW scales
simultaneously will certainly require a deeper investigation which is beyond
the scope of the present paper and will be taken up in near future studies.

\medskip

\section*{Acknowledgements}
The research was supported in part by the European Regional Development Fund
under Grant No.~TK133. The authors thank Alberto Salvio and Florian Nortier
for comments.

\begin{appendix}
\section{Dyson--Schwinger equation for the one-point correlation
function of the Higgs sector}
\setcounter{equation}{0}\def\theequation{A\arabic{equation}}

In order to simplify the computation, we omit the gauge fields. Therefore, for
the equation of motion we obtain
\begin{equation}
\partial^2H+\lambda H|H|^2=j,
\end{equation}
where $j$ is an arbitrary current transforming as an element of
$SU(2)\times U(1)$. Using the partition function 
\begin{equation}
{\cal Z}_H[j^\dagger,j]=\int[dH^\dagger][dH]
e^{i\int\left[|\partial H|^2-\frac{\lambda}{2}|H|^4
-i(j^\dagger H+jH^\dagger)\right]d^4x},
\end{equation}
one has
\begin{equation}
\langle H(x)\rangle={\cal Z}_H[j^\dagger,j]G_1^{(j)}(x).
\end{equation}
Calculating further functional derivatives with respect to the currents
$j$ and $j^\dagger$, one obtains
\begin{eqnarray}
\lefteqn{\strut\kern-12pt{\cal Z}_H[j^\dagger,j]^{-1}\langle|H|^2\rangle
  \ =\ |G_1^{(j)}(x)|^2+G_2^{(j)}(x,x),}\nonumber\\
\lefteqn{\strut\kern-12pt{\cal Z}_H[j^\dagger,j]^{-1}\langle H|H|^2\rangle
  \ =\ G_1^{(j)\dagger}(x)F_2^{(j)}(x,x)}\nonumber\\&&
  +G_2^{(j)}(x,x)G_1^{(j)}(x)+G_1^{(j)\dagger}(x)G_2^{(j)}(x,x)\nonumber\\&&
  +G_1^{(j)}(x)|G_1^{(j)}(x)|^2+G_3^{(j)\dagger}(x,x,x),
\end{eqnarray}
where we have defined
\begin{eqnarray}
G_2^{(j)}(x,y)&=&\frac{\delta^2}{\delta j^\dagger(x)\delta j(y)}
  \ln{\cal Z}_H[j^\dagger,j]=G_2^{\dagger(j)}(y,x),\nonumber\\
F_2^{(j)}(x,y)&=&\frac{\delta^2}{\delta j^\dagger(x)\delta j^\dagger(y)}
  \ln{\cal Z}_H[j^\dagger,j],\nonumber\\
G_3^{(j)}(x,y,z)&=&\frac{\delta}{\delta j^\dagger(z)}G_2^{(j)}(x,y).
\end{eqnarray}
After the Dyson--Schwinger equations have been solved, the currents
$j^\dagger$ and $j$ are put to zero and the upper index $(j)$ of the
$n$-point functions is dropped. 

\section{Dyson--Schwinger equation for the 1P-correlation function
of the Yang--Mills theory}
\setcounter{equation}{0}\def\theequation{B\arabic{equation}}

In order to obtain the Dyson--Schwinger equation for the one-point (1P)
correction function, we assume the existence of a partition function
${\cal Z}[j,\bar\eta,\eta]$ that generates the correlation functions, where
$j_\mu^a$ is a current for the non-Abelian field and $\bar\eta$, $\eta$ are
the currents for the ghost fields. In our case, we have to evaluate
\begin{equation}
G_{n\mu_1\cdots\mu_n}^{(j)a_1\cdots a_n}(x_1,\ldots,x_n)
  =\frac{\delta^n\ln{\cal Z}[j,\bar\eta,\eta]}{\delta j^{a_1}_{\mu_1}(x_1)
  \cdots\delta j^{a_n}_\mu(x_n)}.
\end{equation}
For the ghost field we will have
\begin{equation}
P_n^{(\eta)a_1\cdots a_n}(x_1,\ldots,x_n)=\frac{\delta^n\ln{\cal Z}[j,
  \bar\eta,\eta]}{\delta\bar\eta^{a_1}(x_1)\cdots\delta\eta^{a_n}(x_n)},
\end{equation}
and similarly ${\bar P}_n$ with respect to ${\bar\eta}$, where the functional
derivatives with respect to the currents $\bar\eta$ and $\eta$ are applied
alternately. We will also have correlation functions obtained through mixed
derivatives with respect to $j$ and $\eta$, $\bar\eta$ like $K_n$ and
${\bar K}_n$. In order to obtain the first Dyson--Schwinger equation, we
consider the equations of motion 
\begin{eqnarray}
\lefteqn{\partial^\mu\partial_\mu A^a_\nu+\left(1-\frac1\xi\right)
  \partial_\nu(\partial^\mu A^a_\mu)}\nonumber\\&&
  +gf^{abc}A^{b\mu}(\partial_\mu A^c_\nu-\partial_\nu A^c_\mu)
  +gf^{abc}\partial^\mu(A^b_\mu A^c_\nu)\nonumber\\&&
  +g^2f^{abc}f^{cde}A^{b\mu}A^d_\mu A^e_\nu
  \ =\ gf^{abc}\partial_\nu(\bar c^b c^c)+j^a_\nu,
\end{eqnarray}
where $\xi$ is gauge fixing parameter. Taking the average with respect to the
partition function and using
$\langle A_\mu^a(x)\rangle=G_{1\mu}^a(x){\cal Z}[j,\bar\eta,\eta]$,
$\langle\bar c^b(x)\rangle=\bar P_1^b(x){\cal Z}[j,\bar\eta,\eta]$,
$\langle c^c(x)\rangle=P_1^c(x){\cal Z}[j,\bar\eta,\eta]$ and
\begin{eqnarray}
\lefteqn{{\cal Z}^{-1}\langle A^b_\mu(x)A^c_\nu(x)\rangle
  \ =\ G_{2\mu\nu}^{(j)bc}(x,x)+G_{1\mu}^{(j)b}(x)G_{1\nu}^{(j)c}(x),}
  \nonumber\\
\lefteqn{{\cal Z}^{-1}\langle\partial_\mu A_\nu^c(x)
  -\partial_\nu A_\mu^c(x)\rangle\ =\ \partial_\mu G_{1\nu}^{(j)c}(x)
  -\partial_\nu G_{1\nu}^{(j)}c(x),}\nonumber\\
\lefteqn{{\cal Z}^{-1}\langle A^{\mu b}(x)\left(\partial_\mu
  A_\nu^c(x)-\partial_\nu A_\mu^c(x)\right)}\nonumber\\
  &=&\partial_\mu G_{2\nu}^{(j)\mu bc}(x)+G_1^{(j)\mu b}(x)\partial_\mu
  G_{1\nu}^{(j)c}(x)\nonumber\\&&
  -\partial_\nu G_{2\mu}^{(j)\mu bc}(x,x)
  -G_1^{(j)\mu b}(x)\partial_\nu G_{1\mu}^{(j)c}(x),\nonumber\\
\lefteqn{{\cal Z}^{-1}\langle\partial^\mu\left(A_\mu^b(x)
  A_\nu^c(x)\right)\rangle}\nonumber\\
  &=&\partial^\mu G_{2\mu\nu}^{(j)bc}(x,x)+\partial^\mu(G_{1\nu}^{(j)b}(x)
  G_{1\nu}^{(j)c}(x),\nonumber\\
\lefteqn{{\cal Z}^{-1}\langle A^{\mu b}(x)A_\mu^d(x)
  A_\nu^e(x)\rangle\ =\ G_{3\mu\nu}^{(j)\mu bde}(x,x,x)}\nonumber\\&&
  +G_{2\mu}^{(j)\mu bd}(x,x)G_{1\nu}^{(j)e}(x)+G_{2\nu}^{(j)\mu be}(x,x)
  G_{1\mu}^{(j)d}(x)\nonumber\\&&
  +G_{2\mu\nu}^{(j)de}(x,x)G_{1}^{(j)\mu b}(x)
  +G_{1}^{(j)\mu b}(x)G_{1\mu}^{(j)d}(x)G_{1\nu}^{(j)e}(x),\nonumber\\
\lefteqn{{\cal Z}^{-1}\langle\bar c^b(x)c^c(x)\rangle
  \ =\ P^{(\eta)cb}_2(x,x)+\bar P_1^{(\eta)b}(x)P_1^{(\eta)c}(x),}\nonumber\\
\lefteqn{{\cal Z}^{-1}\langle c^a(x)A_\nu^b(x)\rangle
  \ =\ K_{2\nu}^{(\eta,j)ab}(x,x)+P_1^{(\eta)a}(x)G_{1\nu}^{(j)b}(x)}
  \nonumber\\
\end{eqnarray}
(${\cal Z}^{-1}:={\cal Z}[j,\bar\eta,\eta]^{-1}$), for Feynman gauge $\xi=1$
one obtains the first of the Dyson--Schwinger equations~(\ref{DSE}). The
second one is found in a similar way.

\end{appendix}

\end{document}